\documentstyle{article}
\begin{document}

\vspace{1.5 mm}
INFN-NA-IV-93/20~~~~~~~~~~~~~~~~~~~~~~~~~~~~~~~~~~~~~~DSF-T-93/20
\vspace{2cm}
\begin{center}
{\bf Quantum Limits in Interferometric GW Antennas in the Presence of Even and
Odd coherent states}
\end{center}

\bigskip
\begin{center}
Nadeem A. Ansari, L. Di Fiore, M. A. Man'ko$^{*}$,
V.I. Man'ko$^{*}$, S. Solimeno and F. Zaccaria\\
\end{center}
\bigskip
\begin{center}
Dipartimento di Scienze Fisiche, Universita` di Napoli "Federico II"\\
Istituto Nazionale di Fisica Nucleare, Sezione di Napoli
Mostra d'Oltremare, Pad.20, 80125 Napoli, Italy
\end{center}
\bigskip
\footnote {*~~On leave from Lebedev Physics Institute, Moscow, Russia.}
\newpage
\begin{abstract}

 We discuss a model for interferometric
GW antennas without dissipative or active elements.
 It is predicted that the even and odd
coherent states may play an alternative role to squeezed vacuum states
in reducing the optimal power of the input laser.
\end{abstract}
\newpage

The problem of detecting gravitational waves has been a subject of interest
since last many years [1]. Specifically,
 the quantum sensitivity of Michelson interferometric
gravitational wave detection has been described in detail by Caves [2]. An
important ingredient  improving the sensitivity of such
detectors (GWD) is  using the appropriate states of light beam through the
two ports of the Michelson interferometer. Caves [2] showed, in fact,
 that if one uses coherent light [3] from the first port of the
 interferometer, then the
optimal sensitivity  is limited by the vacuum fluctuations
which enter through the unused port of the interferometer. In such set up
the lower limit on the optimal power of the input laser
 comes out to be quite large and of no experimental interest.
Caves [2] suggested   to reduce
 considerably  the above limit by squeezing the vacuum field entering
through the unused port [4].

The main purpose of this communication is to answer the following question:
"{\it are there any other nonclassical states different from squeezed states,
 which can replace squeezed vacuum in GWD for a better quantum sensitivity
of the Michelson interferometer}"? We predict a possible positive
 answer to this
question in the use of even or odd coherent states [5]. Even coherent states
are
closely related to the squeezed vacuum
states because they too are superposition of
even number states, but with different
coefficients. Different nonclassical properties of even coherent states and
theoretical predictions for their possible generation have been discussed
in detail in Refs.(6-11).

In the Michelson interferometer (Fig.1) we have two incoming fields through
ports $P_{i},~i=1,2$, described by the operators $(a_{i},a_{i}^{\dag})$
acting on a Hilbert space ${\cal H}^{a}={\cal H}_{1}^{a}\otimes
{\cal H}_{2}^{a}$.
To them corresponds two fields at the mirrors $M_{i}$ described by $(b_{i},
b_{i}^{\dag})$ acting on ${\cal H}^{b}={\cal H}_{1}^{b}\otimes{\cal H}_{2}^{b}$
and two outgoing fields at $P_{i}$ described by $(c_{i},c_{i}^{\dag})$ on
${\cal H}^{c}={\cal H}_{1}^{c}\otimes{\cal H}_{2}^{c}$. The basis in
${\cal H}_{i}^{\rho},~\rho=a,b,c$, will be denoted $\{\mid n,i,\rho>,~~~
n\in Z^{+}\}$.
We simplify the Michelson
interferometer as a device with two arms at the end of which two outer
mirrors $M_{i}$ are attached to some string, thus behaving as two pendula,
without considering Fabry-Perot cavities and beam delaying optics into these
arms.
We will suppose that in all processes the dissipative and
active effects are negligible
so that conservation of energy in ensured. The Hamiltonian in ${\cal H}^{\rho}$
is taken to be
\begin{equation}
H^{\rho}=\hbar \omega (\rho_{1}^{\dag}\rho_{1}+\rho_{2}^{\dag}\rho_{2})
\end{equation}
with $\omega$ the frequency and $\hbar$ the Planck's constant. Implicit here
is the assumption of equal frequencies for the mode 1 and 2. This can
be achieved by introducing some degree of interaction among the two
modes, which anyhow can be ignored in a first approximation, as in Ref.[2].
All ${\cal H}^{\rho}$ are unitarily equivalent and the operators $H^{\rho}$
are connected each other by 2$\times$ 2 unitary matrices,
the elements of which depend on the physical and geometrical parameters
of the interferometer. For instance, we will write
\begin{equation}
b=V~a,~~~~b^{\dag}=a^{\dag}~V^{\dag}
\end{equation}
where
\begin{eqnarray}
a&=&\left( \begin{array}{c}
a_{1}\\a_{2}
\end{array}\right)~;~~b=\left( \begin{array}{c}
b_{1} \\ b_{2} \end{array} \right) \nonumber\\
a^{\dag}&=&(a_{1}^{\dag}~~a_{2}^{\dag})~~;~b^{\dag}=(b_{1}^{\dag}~~b_{2}^{\dag})
\end{eqnarray}
and $V\in U(2)$ group. We conveniently write
\begin{equation}
V=\Phi K\nonumber\\
\end{equation}
with
\begin{eqnarray}
\Phi&=&\left( \begin{array}{clcr}
e^{i\phi_{1}} & 0 \\
0 & e^{i \phi_{2}} \end{array} \right)\nonumber\\
K&=&\left( \begin{array}{clcr}
\alpha_{1} & \beta_{2}\\
\beta_{1} & \alpha_{2} \end{array} \right)
\end{eqnarray}
In the above $\alpha_{i}$ and $\beta_{i}$ are the complex transmittivity and
reflectivity parameters of the beam splitter (BS)  arbitrarily oriented
for the ith input field mode respectively
and $\phi_{i}$ is the phase distance between
BS and the mirror $M_{i}$. Also
\begin{equation}
c=U~a~;~~c^{\dag}=a^{\dag}U^{\dag}
\end{equation}
and
\begin{equation}
U=-K^{T}\Phi^{2} K=-V^{T}V
\end{equation}
The presence of negative sign in the above equation is due to the phase
change on reflections at the mirrors.

Following [2], we have two sources of errors which set the lower quantum
limit $\Delta z$ on the sensitivity z of GW antennas: (i) radiation pressure
 on $M_{i}$ and (ii)
 photon counting noise due to the fluctuations in the
number of photons in the input fields
\begin{equation}
\Delta z=\sqrt{(\Delta z_{rp})^{2}+(\Delta z_{pc})^{2}}
\end{equation}
In this
\begin{equation}
(\Delta z_{rp})^{2} = \sigma_{rp}^{2}(\hbar \omega \tau/mc)^{2}
\end{equation}
where
\begin{equation}
\sigma_{rp}^{2}=<(b^{\dag}\sigma_{3}b)^{2}>-<b^{\dag}\sigma_{3}b>^{2}
\end{equation}
and
\begin{equation}
(\Delta z_{pc})^{2} = \sigma_{pc}^{2}~\left( \frac{\partial
<c^{\dag}\sigma_{3}c>}
{\partial (\phi_{2}-\phi_{1})} \right)^{-2}
\end{equation}
where
\begin{equation}
\sigma_{pc}^{2}=<(c^{\dag}\sigma_{3}c)^{2}>-<c^{\dag}\sigma_{3}c>^{2}
\end{equation}
In the above, $\tau$ is the observation  time and m  the mass of
each end mirrors.
Here we consider fixed  BS as in Ref.(2).
After little algebra we can write
\begin{eqnarray}
\sigma_{rp}^{2}&=&(U^{\dag}\sigma_{3}U)_{ik}(U^{\dag}\sigma_{3}U)_{mn} T_{ikmn}
\nonumber\\
\sigma_{pc}^{2}&=&(V^{\dag}\sigma_{3}V)_{ik}(V^{\dag}\sigma_{3}V)_{mn} T_{ikmn}
\end{eqnarray}
with the summation over repeated indices  taken from 1 to 2 and
\begin{equation}
T_{ikmn}=<a_{i}^{\dag}a_{k}a_{m}^{\dag}a_{n}>-<a_{i}^{\dag}a_{k}>
<a_{m}^{\dag}a_{n}>
\end{equation}
 This allows an easy comparison between situations arising from the
use of different types of input fields.

Combining Eqs.(8-13) yields
\begin{equation}
(\Delta z)^{2}=X_{ikmn}~T_{ikmn}~~~~(ikmn=1,2)
\end{equation}
where $X_{ikmn}$ contain the geometrical and physical properties of the
antenna while the second factors $T_{ikmn}$  depend only on  the
incoming fields.

For a simple Michelson interferometer,
Caves suggested to use squeezed vacuum light in order to minimize the input
laser power [2]. Equation (15) permits to investigate
very general states of the input field like even and odd coherent states,
 correlated states, states with higher order squeezing, etc. In particular,
we will illustrate in this communication the dependence of the optimal
$\Delta z$ on the characteristic parameters of the even or odd coherent
states from the second port of interferometer.\\

 First, we will evaluate the matrix $X_{ikmn}$ as far as the geometrical and
physical parameters of the Michelson interferometer are concerned.

If we consider a
50-50 \% BS, then the elements of the matrix  $K$ are
\begin{eqnarray}
\alpha_{1}=\alpha_{2}&=&\frac{e^{i\delta}}{\sqrt{2}}\nonumber\\
\beta_{1}=\beta_{2}&=&\frac{e^{i\mu}}{\sqrt{2}}
\end{eqnarray}
where $\delta$ is the phase because of the BS which can be set to zero
for an ideally thin
BS while $\mu$ is the phase introduced by the BS between
the reflected and transmitted waves and for simplicity we  take $\mu=\pi /2$.
Then
\begin{equation}
V^{\dag}\sigma_{3}V=\left( \begin{array}{clcr}
0 & i \\
-i & 0 \end{array} \right)
\end{equation}
and
\begin{equation}
U^{\dag}\sigma_{3}U=\left ( \begin{array}{clcr}
-\cos\phi & -\sin\phi \\
-\sin\phi & \cos\phi \end{array} \right)
\end{equation}
in $\phi=2(\phi_{2}-\phi_{1})$. Also
\begin{equation}
<c^{\dag}\sigma_{3}c>=<a^{\dag}\sigma_{3}a>\cos\phi
\end{equation}
If the interferometer has to operate in a dark fringe then the arms lengths
can be adjusted to have $\phi=\frac{(2n+1)\pi}{2}$ and  dark
fringes correspond to the situation where  no field contributions is present
into the difference of the output photon numbers. In such cases we have
\begin{equation}
U^{\dag}\sigma_{3}U=\left ( \begin{array}{clcr}
0 & -1 \\
-1 & 0 \end{array} \right)
\end{equation}
 Then $X_{ikmn}$ become
\begin{eqnarray}
X_{1212}=X_{2121}&=&-A^{2}+B^{2}\nonumber\\
X_{1221}=X_{2112}&=&A^{2}+B^{2}
\end{eqnarray}
where
\begin{eqnarray}
A&=& \left( \frac{\hbar \omega \tau}{mc} \right)\nonumber\\
B&=&\left( \frac{\partial I}{ \partial Z} \right)^{(-1)}
\end{eqnarray}
and
\begin{eqnarray}
I&=&<c^{\dag}\sigma_{3}c>\nonumber\\
Z&=&\phi \frac{c}{2\omega}
\end{eqnarray}
The variable Z corresponds to the difference between the displacements of the
two outer mirrors caused by the radiation pressure
with respect to their mean positions in the absence of any field.

We will now evaluate the factors $T_{ikmn}$ in (i) Caves setup  and (ii)
in a new one
which replaces the squeezed light with even or odd coherent light.

In order to evaluate the contribution of the fields which are
applied to the two ports of the interferometer for GW detection,
we will assume a coherent light for the field relative to port one of the
interferometer while for the second port we will consider the two situations
(i) by squeezing the vacuum fluctuations
 (the situation considered by the Caves[2]) and
(ii) applying even or odd coherent states. We will
 show the important role played by these states
in order to get a better detection sensitivity and to reduce the optimal
input laser power.

When
coherent laser light from port one and squeezed vacuum from the other port
of the interferometer are applied, the two fields are anticorrelated.
The states of ${\cal H}^{a}$ can be written as
\begin{equation}
\mid \psi>={\cal D}_{1}(\alpha)\mid 0,1,a>
e^{\frac{\xi a_{2}^{\dag 2}-\xi^{*} a_{2}^{2}}{2}}\mid 0,2,a>
\end{equation}
where ${\cal D}_{i}(\alpha)=e^{(\alpha a_{i}^{\dag}-\alpha^{*}a_{i})}$
 i=1,2,
$\alpha \in C$, and $\xi=re^{i\theta_{1}}$.
It is easy to see that in such states $<a_{1}a_{2}>$,$<a_{1}^{\dag}a_{2}>$,
etc., are equal to zero.

If we take $\alpha$ to be real for simplicity then we have
\begin{eqnarray}
T_{1111}&=&\alpha^{2}\nonumber\\
T_{1122}&=&0\nonumber\\
T_{1212}&=&-\alpha^{2}sinhr~coshr~e^{i\theta_{1}}\nonumber\\
T_{1221}&=&\alpha^{2}sinh^{2}~r+\alpha^{2}\nonumber\\
T_{2112}&=&\alpha^{2}sinh^{2}~r+sinh^{2}~r\nonumber\\
T_{2121}&=&-\alpha^{2}sinhr~coshr~e^{-i\theta_{1}}\nonumber\\
T_{2211}&=&0\nonumber\\
T_{2222}&=&2sinh^{2}~r
\end{eqnarray}

When even or odd coherent states replace squeezed vacuum in port two
the states of ${\cal H}^{a}$ to be taken into account are
\begin{eqnarray}
\mid \psi>&=&\mid \alpha,\beta_{\pm}>\nonumber\\
&=&{\cal D}_{1}(\alpha)\mid 0,1,a>
N_{\pm}({\cal D}_{2}(\beta)\pm {\cal D}_{2}(-\beta) \mid 0,2,a>
\end{eqnarray}
where `+,-' signs correspond to even and odd coherent states respectively
and their normalization constants are
\begin{eqnarray}
N_{+}&=&\frac{1}{2e^{-\frac{\mid \beta \mid^{2}}{2}}\sqrt{cosh
\mid \beta \mid^{2}}}\nonumber\\
N_{-}&=&\frac{1}{2e^{-\frac{\mid \beta \mid^{2}}{2}}\sqrt{sinh
\mid \beta \mid^{2}}}
\end{eqnarray}
For the even light, coefficients $T_{ikmn}$ take the following values
\begin{eqnarray}
T_{1111}&=&\alpha^{2}\nonumber\\
T_{1122}&=&0\nonumber\\
T_{1212}&=&\alpha^{2}\mid \beta \mid^{2}e^{2i\theta_{2}}\nonumber\\
T_{1221}&=&\alpha^{2}\mid \beta \mid^{2}tanh\mid \beta \mid^{2}
+\alpha^{2}\nonumber\\
T_{2112}&=&\alpha^{2}\mid \beta \mid^{2}tanh\mid \beta \mid^{2}+
\mid \beta \mid^{2}tanh\mid \beta \mid^{2}\nonumber\\
T_{2121}&=&\alpha^{2}\mid \beta \mid^{2}e^{-2i\theta_{2}}\nonumber\\
T_{2211}&=&0\nonumber\\
T_{2222}&=&\mid \beta \mid^{4}-\mid \beta \mid^{4}tanh^{2}~\mid \beta \mid^{2}
+\mid \beta \mid^{2}tanh\mid \beta \mid^{2}
\end{eqnarray}
in which $\theta_{2}$ is the phase of  $\beta$.

For odd coherent states we get the same expressions as above, except that
$tanh\mid \beta \mid^{2}$ should be replaced by $coth\mid \beta \mid^{2}$.

The general expression for $(\Delta z)^{2}$, irrespective of the nature of the
incoming fields can now be written as
\begin{equation}
(\Delta z^{2})=A^{2}(T_{1221}+T_{2112}-T_{1212}-T_{2121})
+B^{2}(T_{1221}+T_{2112}+T_{1212}+T_{2121})\nonumber\\
\end{equation}

This quantity depends on the incoming field through $P_{1}$ and we denot by
 $(\alpha_{opt}^{2})^{(o)}$
=$m c^{2}/(2 \hbar \omega \tau)$ the intensity of this field which minimizes
$(\Delta z)^{2}$ when at $P_{2}$ the ordinary vacuum is present. Then, it can
be seen that the value which minimizes $(\Delta z)^{2}$, which we call
$(\alpha_{opt}^{2})^{(sq)}$, in presence of
squeezed vacuum at $P_{2}$, under the condition
$\alpha^{2}>>sinh^{2}~r$ and $\theta_{1}=0$, is
\begin{equation}
(\alpha_{opt}^{2})^{(sq)}= e^{-2r}~(\alpha_{opt}^{2})^{(o)}
\end{equation}
This is the Caves result, which allows to reduce the intensity of the input
laser beam to values experimentally significant.

The analogous analysis for the cases of even or odd coherent states replacing
squeezed vacuum, under the condition that
$\alpha^{2}>>\mid \beta \mid^{2}tanh \mid \beta \mid^{2}$ gives
\begin{equation}
(\alpha_{opt}^{2})^{(ev)}=
\sqrt{\frac{ 2\mid \beta \mid^{2}tanh \mid \beta \mid^{2}+2\mid \beta \mid^{2}
cos~2\theta+1}{ 2\mid \beta \mid^{2}tanh \mid \beta \mid^{2}-
2\mid \beta \mid^{2}cos~2\theta+1}}(\alpha_{opt}^{2})^{(o)}
\nonumber\\
\end{equation}
and
\begin{equation}
(\alpha_{opt}^{2})^{(odd)}=
\sqrt{\frac{ 2\mid \beta \mid^{2}coth \mid \beta \mid^{2}+2\mid \beta \mid^{2}
cos~2\theta+1 }{ 2\mid \beta \mid^{2}coth \mid \beta \mid^{2}-
2\mid \beta \mid^{2}cos~2\theta+1}}(\alpha_{opt}^{2})^{(o)}
\nonumber\\
\end{equation}
Thus by using even coherent light is used, under the limit
$1\ll \mid \beta \mid^{2} \ll \alpha^{2}$ and $\theta_{2}=\pi/2$, yields
\begin{equation}
(\alpha_{opt}^{2})^{(ev)}=\frac{(\alpha_{opt}^{2})^{(o)}}{2 \mid \beta \mid }
\end{equation}
This result, which is also true for odd coherent states,
 allows an alternative way to decrease  the optimal input power,
i.e.,
an alternative way to increase the sensitivity of the interferometer. We have,
therefore, given a positive answer to the question originally posed.
The question whether this new way might be experimentally achievable or not
is left open, depending on the actual physical generation  of the even and odd
coherent states.

We wish now to consider  more general situations: namely for case (i)
$\xi$  is arbitrary and for case (ii) $\beta$ is arbitrary
and in both situations $\alpha$ is real. $(\Delta z)^{2}$ is a function
of such variables and we can look for its minimization with respect to
$\alpha^{2}$. This results for case (i) in $\theta_{1}=0$ and
$(\alpha_{opt}^{2})^{(sq)}$ function of r and for case (ii) in
$\theta_{2}=\pi/2$ and
$(\alpha_{opt}^{2})^{(ev)}$, $(\alpha_{opt}^{2})^{(odd)}$  functions
of $\mid \beta \mid$. In each case, $\alpha_{opt}^{2}$ as function of the
respective independent variable, is given through a solution of the following
equation in $\xi$
\begin{equation}
\Gamma_{1} \xi^{3}-\Gamma_{2} \xi -2\Gamma_{3} (1+\Gamma_{2})=0
\end{equation}
in which
\begin{equation}
\xi=\frac{\alpha_{opt}^{2}}{(\alpha_{opt}^{2})^{(o)}}-\Gamma_{3}
\end{equation}
and
\begin{eqnarray}
\Gamma_{1} &=&e^{2r}\nonumber\\
\Gamma_{2}&=&e^{-2r}\nonumber\\
\Gamma_{3} &=& \frac{sinh^{2}~r}{(\alpha_{opt}^{2})^{(o)}}
\end{eqnarray}
for the squeezed vacuum, and
\begin{eqnarray}
\Gamma_{1} &=&2\mid \beta \mid^{2}tanh\mid \beta \mid^{2}-2\mid \beta \mid^{2}
cos2\theta_{2}+1\nonumber\\
\Gamma_{2} &=&2\mid \beta \mid^{2}tanh\mid \beta \mid^{2}+2\mid \beta \mid^{2}
cos2\theta_{2}+1\nonumber\\
\Gamma_{3} &=& \frac{\mid \beta \mid^{2}tanh\mid \beta \mid^{2}}
{(\alpha_{opt}^{2})^{(o)}}
\end{eqnarray}
for the even coherent states. For the quantities relative to odd coherent
states
same formulas apply with  $coth \mid \beta \mid^{2}$ in place of
$tanh \mid \beta \mid^{2}$.

Equation (34) has three roots, two of them complex and the physical one
has the following form
\begin{eqnarray}
\alpha^{\prime} \equiv \frac{\alpha_{opt}^{2}}{(\alpha_{opt}^{2})^{o}}&=&
\frac{3^{\frac{3}{2}} \sqrt{\Gamma_{1}} \Gamma_{4}
 \Gamma_{3}+9 \Gamma_{2}+\Gamma_{4}^{2}}
{3^{\frac{3}{2}}\sqrt{\Gamma_{1}}\Gamma_{4}}\nonumber\\
\Gamma_{4}&=&(9\Gamma_{3} \sqrt{\Gamma_{1}}(1+\Gamma_{2})+\sqrt{81\Gamma_{1}
\Gamma_{3}^{2}(1+
\Gamma_{2})^{2}-\Gamma_{2}^{3}})^{\frac{1}{3}}
\end{eqnarray}
 Fig.(2), illustrates case (i) and $\alpha^{\prime}$ is plotted versus the
squeezing parameter r. For r=0 we have  the situation in which the only vacuum
fluctuations enter from $P_{2}$
and in this case $\alpha^{\prime}$=1. For large values of r,
we have Caves result [2], i.e., the optimal value of the input coherent field
through $P_{1}$
can be decreased to a large amount.

 With the same spirit, in Fig.(3a), we have plotted
$(\alpha^{\prime})_{ev}^{-1}$ versus $\mid \beta \mid$ and $\theta_{2}$
and in Fig.(3b) this quantity versus $\mid \beta \mid$ at $\theta_{2}=\pi/2$,
which is the value for all the local minima, when even coherent light enters
from port 2. The analogous graphical analysis for odd coherent states
is shown in Figs.(4a) and (4b).

Such figures illustrate how the optimal values of the input coherent field
in $P_{1}$ can be reduced considerably and allows us to predict an
application of  even or odd coherent fields.

In conclusion it has been shown that such states might offer
a new technique to reduce the optimal power of the input coherent
laser and for a better sensitivity of the interferometer.

\newpage

{\bf Acknowledgements}\\

M.A.M. and V.I.M. wish to thank the Dipartimento di Scienze Fisiche, of the
Universita`
di Napoli and I.N.F.N. for the kind hospitality. The research of N.A.A. was
supported by the International Center for Theoretical Physics Programme for
Research and Training in Italian laboratories, Trieste, Italy.

\bigskip

{\bf References}\\

\begin{enumerate}

\item see for example, "The detection of gravitational waves", edited
by  D.G. Blair (Cambridge University Press, Cambridge, 1991),
and the references there in.

\item C.M. Caves, Phys. Rev. A {\bf 23}, (1981) 1693; Phys. Rev. Lett.
{\bf 45}, 75 (1980).

\item R.J. Glauber, Phys. Rev. Lett. {\bf 10}, 84 (1963).

\item D. Stoler, Phys. Rev. D {\bf 4} 2309 (1971); J.N Hollenhorst Phys. Rev. D
{\bf 19}, 1669 (1979); E.Y. Lu, Lett. Nuovo Cim. {\bf 3}, 585 (1972);
D.P. Bertrand, K. Moy. E.A. Mishkin, Phys. Rev D {\bf 4}, 1909 (1971);
H.P. Yuen, Phys. Rev A {\bf 13}, 2226 (1976); D.F. Walls Nature {\bf 306},
141 (1983).

\item  V. V. Dodonov, I. A. Malkin, and V. I. Man'ko, Physica {\bf 72}, 597
(1974).

\item J. Gea-Banacloche, Phys. Rev. A {\bf 44}, 5913 (1991).

\item  C. C. Gerry and E. E. Hach III, Phys. Lett. {\bf A 74},
185 (1993)

\item V. Buzek, A. Vidiella-Barranco and P.L. Knight, Phys. Rev. A {\bf 45},
6570 (1992).

\item J. Jansky and A.V. Vinogardov, Phys. Rev. Lett. {\bf 64}, 2771 (1990).

\item J. Perina, Quantum Statistics of Linear and Non-Linear Optical
Phenomena (Reidel, Dordrecht, 1984) p. 78.

\item C.C. Gerry Optics Commun. {\bf 91}, 247 (1992).

\item S. Solimeno, F. Barone, C.de Lisio, L.Di Fiore, L. Milano, and G. Russo,
Phys. Rev. A {\bf 43}, (1991) 6227.

\end{enumerate}

\newpage

{\bf Figure Captions}\\

\begin{description}

\item[Fig.(1).] Schemetic of the simple Michelson interferometer $a_{1},~a_{2}$
and  , $c_{1},~c_{2}$ are respectively the input and output fields, while
$b_{1},~b_{2}$
stand for the fields incident on mirrors , $M_{1}$ and $M_{2}$.

\item[Fig.(2).] Relative value $(\alpha^{\prime})_{sq}$ of the optimal
laser intensity in the presence and of the squeezed vacuum mode $a_{2}$, versus
the squeezing parameter r.

\item[Fig.(3a).] Three dimensional plot of the relative value
$(\alpha^{\prime})_{ev}^{-1}$ of the optimal laser intensity
in the presence of even coherent states  versus
$\mid \beta \mid$ and $\theta_{2}$.

\item[Fig.(3b).] $(\alpha^{\prime})_{ev}$ versus $\mid \beta \mid$ for
$\theta_{2}=\pi/2$.

\item[Fig.(4a)] Three dimensional plot of the relative value
$(\alpha^{\prime})_{odd}^{-1}$ of the optimal laser intensity
in the presence and in the absence of  odd coherent states  versus
$\mid \beta \mid$ and $\theta_{2}$.

\item[Fig.(4b).]  $(\alpha^{\prime})_{odd}$ versus $\mid \beta \mid$ for
$\theta_{2}=\pi/2$.

\end{description}

\end{document}